# DCA: Dynamic Correlation Analysis

Tianwei Yu

Department of Biostatistics and Bioinformatics, Emory University, Atlanta, GA 30322, USA.  Email: tianwei.yu@emory.edu.

**Abstract**

In high-throughput data, dynamic correlation between genes, i.e. changing correlation patterns under different biological conditions, can reveal important regulatory mechanisms. Given the complex nature of dynamic correlation, and the underlying conditions for dynamic correlation may not manifest into clinical observations, it is difficult to recover such signal from the data. Current methods seek underlying conditions for dynamic correlation by using certain observed genes as surrogates, which may not faithfully represent true latent conditions. In this study we develop a new method that directly identifies strong latent signals that regulate the dynamic correlation of many pairs of genes, named DCA: Dynamic Correlation Analysis. At the center of the method is a new metric for the identification of gene pairs that are highly likely to be dynamically correlated, without knowing the underlying conditions of the dynamic correlation. We validate the performance of the method with extensive simulations. In real data analysis, the method reveals novel latent factors with clear biological meaning, bringing new insights into the data.

## Introduction

The cellular system involves tens of thousands of genes/proteins that are tightly regulated in a complex network (1-3). Interactions and regulations in the network are highly dynamic. They change substantially in different cell types, developmental stages, or in response to environmental conditions (4). Gene expression and similar types of data, such as proteomics and metabolomics data, represent outcomes of the dynamic regulatory network. Changes in the underlying regulation patterns are reflected in the changes in gene expression levels, and/or changes in the correlation between genes. Many methods are available to analyze patterns in the gene expression levels (5-8), while less attention has been paid to the study of dynamic correlations.

Methods have been developed to find differential correlation patterns between genes or gene sets, conditioned on a given clinical variable (9-11). However, dynamic correlation can be more complex. Underlying cellular states may not manifest into clinical observations. As the biological system is regulated in a modular manner (12), there could be multiple dynamic correlation conditions that govern different functional groups of genes. Hence it is of interest to find unobserved dynamic correlation conditions, which is a much harder problem. To this end, Li has developed the Liquid Association (LA) approach, which uses a third gene as the proxy of the dynamic correlation signal (13, 14). The method scans through all possible gene triplets to find potential dynamic correlations. Similar approaches that utilize genes are mediators (15, 16), integrative analysis utilizing LA (17, 18), as well as some statistical theory of LA (19) were later developed.

Although focusing on gene-level dynamic correlations can reveal some important local regulatory mechanisms, a more global approach to dynamic correlation could discover critical regulation mechanisms that penetrate multiple biological processes, or help identify hidden sub-groups in the samples. To this end, using the original LA or similar approaches is not effective due to the following reasons. First, scanning through all possible triplets is computationally intensive. Second, a genome-scale scan yields large numbers of LA gene triplets, causing difficulties in the interpretation. Given the LA score is calculated in a symmetric manner among the three genes involved, discerning which gene reflects cellular states could be tricky. Third and the most important, the genes that serve as surrogate variables may not be good indicators of true underlying cellular states.

In this study, our purpose is to find dominant dynamic correlation signals that regulate the dynamic correlation of a large number of gene pairs. The biggest difficulty is we do not know *a priori* which gene pairs have the relationship of dynamic correlation. We design a new metric, named Liquid Association Coefficient (LAC), to effectively and efficiently screen all gene pairs for potential dynamic correlations. From gene pairs that are most likely to be dynamically correlated, we provide a simple and straight-forward solution for quickly finding the latent dynamic correlation signals. The procedure is named DCA: Dynamic Correlation Analysis. We refer to the latent signals found by DCA as Dynamic Components (DCs).

We demonstrate the performance of the method using extensive simulations. In real biological datasets, we demonstrate the method can identify latent signals that are biologically meaningful and not found by existing methods. In a merged cell cycle dataset, the method can find signals pertaining to the original experimental grouping, as well as biological processes that differentiate between the experiments. In the TCGA breast cancer (BRCA) dataset, the new method can find new interesting subgroups in the subjects that are related to patient survival outcome.

**Methods**

*The overall framework*

The data is in the form of an expression matrix, $\boldsymbol{G}_{p \times n}$, with *p* genes in the rows and *n* samples in the columns. Our assumption is that a portion of the gene pairs have dynamic correlations, and there are some major latent signals that can explain much of the variation in correlations among those gene pairs. Our purpose is to detect such dynamic correlation signals.

We assume that all genes are normalized to have mean 0 and standard deviation 1. Thus the covariance and correlation between two genes $X$ and $Y$ are equal to $E(XY)$. First we assume we know which $m$ gene pairs have the relationship of dynamic correlation. We address the selection of such gene pairs in the next sub-section. Given these gene pairs, we can construct a new matrix $\boldsymbol{B}_{m \times n}$, in which the each row is constructed by multiplying the corresponding elements of a gene pair $X$ and $Y$, $(x_1 y_1, x_2 y_2, \ldots, x_n y_n)$. A gene can contribute to multiple rows of the $\boldsymbol{B}$ matrix if it has dynamic correlation with multiple genes.

For any $\boldsymbol{z}$ vector that is normally distributed, $\boldsymbol{Bz} = (LA_1, LA_2, \ldots, LA_m)'$ is proportional to the LA scores with $\boldsymbol{z}$ being the LA scouting gene over all the pairs. From a clustering perspective, if we find clusters of rows in the matrix $\boldsymbol{B}$, then each cluster shares a common LA scouting factor. Alternatively, from a principal component perspective, $(\boldsymbol{Bz})'(\boldsymbol{Bz})/\boldsymbol{z}'\boldsymbol{z}$ is proportional to the sum of LA scores squared over all the gene pairs. Finding a sequence of unit vectors $\boldsymbol{z}$ that are orthogonal to each other and maximizes the sum of LA scores squared requires the exact same solution as conducting eigenvalue decomposition on the matrix $\boldsymbol{B}'\boldsymbol{B}$.

Conceptually, other methods used to find latent factors, such as Independent Component Analysis (ICA) (20), Sparse Principal Component Analysis (SPCA) (21), Modular Latent Structure Analysis (MLSA) (22), or various clustering methods can also be applied to the $\boldsymbol{B}$ matrix. In this manuscript we focus on the eigenvalue decomposition approach. We note there is a caveat that this approach doesn't guarantee that elements of $\boldsymbol{z}$ will follow the normal distribution.

*Selecting informative gene pairs*

For the purpose of selecting informative gene pairs to find underlying dynamic correlation signals, we define a measure for dynamic correlation between a pair of genes with an unknown condition factor, the Liquid Association Coefficient (LAC), which is the correlation coefficient of the squared values of the two genes, minus the correlation coefficient of the original values squared.

$$\zeta_{i,j} = r(g_i^2, g_j^2) - r^2(g_i, g_j),$$

where $r()$ is the Pearson's correlation coefficient. It has been shown that when both $g_i$ and $g_j$ follow the bivariate normal distribution with mean $\begin{pmatrix}0\\0\end{pmatrix}$, and variance-covariance matrix $\begin{pmatrix}1 & \rho^2\\ \rho^2 & 1\end{pmatrix}$, the population correlation coefficient between $g_i^2$ and $g_j^2$ is equal to $\rho^2$, which makes the above quantity zero.

Alternatively, to reduce the impact of more extreme values, we can use the correlation coefficient of the absolute values of the two genes minus the absolute value of the correlation coefficient:

$$\zeta_{i,j} = r(|g_i|, |g_j|) - |r(g_i, g_j)|.$$

We compute the matrix of *LAC* values for all pairs of genes. Notice the computational cost is on the same scale as computing the pairwise correlation matrix. We then select the $(i,j)$ pairs whose *LAC* values are above a certain percentile of all the values in the matrix.

After selecting the top $(i,j)$ pairs, we construct the **B** matrix, in which each row is constructed from a selected pair of genes. For example, if $g_i$ and $g_j$ are selected as a pair of informative genes, then the corresponding row of the new matrix is $(g_{i1}g_{j1},\ g_{i2}g_{j2}, \ldots,\ g_{in}g_{jn})$. In this study, we use eigenvalue decomposition of **B'B** to extract latent factors, and varimax rotation (23) to improve the interpretability of the latent factors.

*Selecting gene pairs associated with a latent factor*

We first calculate the LAC coefficients for all pairs of genes, and select gene pairs with *LAC* coefficients belonging to a top percentile (20% in this study). We then calculate their LA scores with the latent factor. Heuristically, we model the distribution of LA scores as a mixture, with a dominant split-normal component in the center representing gene pairs with no relation to the latent factor, i.e. the null distribution. We apply the local false discovery (fdr) approach to calculate the posterior probability that a gene pair belongs to the non-null distribution (24), and threshold the fdr values to select gene pairs that are dynamically correlated given the latent factor.

*Finding biological processes associated with a latent factor*

For functional interpretation, we use gene ontology (GO) biological processes. We first select a set of representative GO biological process terms that are of reasonable size and relatively small overlaps, following an existing procedure that considers both the ontology structure and the number of genes assigned to each term (25). For the yeast data, we select 172 biological processes with 50~1000 assigned genes each, covering 5334 genes in total. For the human data, we select 423 biological processes with 100~1000 assigned genes each, covering 14414 genes in total. From the gene pairs associated with each latent factor, we conduct two types of analyses:

**Within-process dynamic correlation.** For each biological process, we count the occurrence of gene pairs in which both genes fall into the process. We also calculate the expected number of such gene pairs if all the gene pairs were randomly drawn. We calculate the fold-change by taking the ratio of observed count *v.s.* the expected count, and p-value using the binomial distribution.

**Between-process dynamic correlation.** For each pair of selected biological processes, we first remove their overlapping genes. We then count the occurrence of gene pairs in which the two genes fall into the two processes respectively, and calculate the expected number of such gene pairs if all the genes were randomly drawn. After thresholding the fold change and p-value to select pairs of processes, we visualize the resulting network using Cytoscape (26).

**Results and Discussion**

*Illustration of the Liquid Association Coefficient (LAC)*

In this study a new metric is defined to rank all pairs of variables in the data matrix. The purpose of the LAC is to help identify gene pairs that are most likely to have the relationship of dynamic correlation, without knowing the underlying conditions of the dynamic correlation. Gene pairs with such relations should receive high LAC score, while other gene pairs, either independent or correlated, should receive low scores.

The LAC requires all variables to have mean zero and standard deviation 1. As illustrated in Figure 1, if both variables X and Y follow the standard normal distribution marginally, and one-third of the (X,Y) pairs are positively correlated, one-third of the (X,Y) pairs are negatively correlated, and another one-third uncorrelated, then the absolute values will be positively correlated, and the *LAC* tends to be large (Fig. 1, left column). On the other hand, when X and Y are truly independent or simply correlated, the *LAC* tends to be small.

We further conduct a larger simulation study to examine the empirical distribution of LAC under different circumstances. As illustrated in Figure 2, when the two variables are dynamically correlated, the distribution of the LAC score is centered at a positive value (Fig. 2, blue curves). The higher the correlation level, the higher the mean (Fig. 2, left to right panels). The higher the sample size, the less the spread (Fig. 2, different line types). At the same time, in the independent and correlated cases, the LAC scores are centered around zero if the first definition of *LAC* is used. Using the second definition, the *LAC* is still centered around zero in the independent case, and the center is negative in the correlated case (Fig. 2, lower panels).

We conducted an extensive simulation study to evaluate the method's capability to recover latent dynamic correlation signals. Please refer to the Supporting Information, section 1 for details (Supporting Figures 1~3). Overall, the method can recover the hidden dynamic correlation signal when the sample size and signal strength is sufficient.

*DCA extracts signals that differentiate experiments from the merged cell cycle data*

We first analyze the well-studied Spellman cell cycle gene expression data (27). The dataset has been analyzed by many authors. The purpose of the analysis here is to demonstrate that DCA can extract information that is clearly meaningful, and provides novel biological insights.

The cell cycle dataset includes four time-series experiments of the yeast cell cycle, each using a different method of synchronization. The total dimension is 6178 genes by 73 samples. Missing values were imputed by the K-nearest neighbor (KNN) method (28). When all four time series datasets are combined into a single dataset, traditional methods such as PCA and SPCA (21) extract signals that are consistent across the four time series (Supporting Figures 4 and 5), but not signals that separate the four time series, except the first PC that captures an oscillating signal which is an artifact in the CDC15 time series data (29).

Applying DCA to the combined cell cycle data yields factors that are distinctly different. Most of the Dynamic Components (DCs) clearly differentiate one of the four time series from the rest (Supporting Figure 6). For a full list of factor plots and biological processes associated with each factor, please refer to Supporting File 2. Here we focus our discussion on three of the factors.

The first DC has high scores for samples from the CDC15 experiment only. It has been documented that an oscillating signal is present in the CDC15 data across many genes (Supporting Figure 7), causing an elevated level of correlation overall (29). The first DC reflects this signal. At the same time, gene pairs associated with this DC are not clearly associated with any biological function, as reflected in the fact that no biological function pairs were found at the threshold of p=0.001 and fold change=2. This is expected given the fact that the oscillating signal is not biologically meaningful.

The second DC only has extreme scores for some of the samples of the elutriation experiment. A closer examination reveals the DC shows a sine-wave pattern in the elutriation samples (Figure 3). An examination of the data reveals a strong dynamic correlation pattern between genes associated with this DC (Supporting Figure 8). Selecting biological processes pairs that have excessive dynamic correlation links between them, we find that the processes are focused on rRNA biogenesis and ribosome assembly. Much more positive/negative correlations are shown between genes in these biological processes when the DC2 score is low, which correspond to half of the samples in the elutriation experiment (Supporting Figure 8). While all the other three experiments are based on block-and-release cell cycle synchronization, the elutriation process separates synchronized cells based on their size, shape and mass (30). The results here indicate that protein biosynthesis tend to be better synchronized in the elutriation samples compared to the other three experiments.

For the fifth DC, samples in the CDC28 experiment have lower scores, while the alpha factor samples have higher scores, with a smaller magnitude (Figure 3). This indicates that some gene pairs have a reverse correlation pattern between the two experiments, which is intriguing given both experiments used block-and-release to synchronize cells. Some more recent studies have shed light on the metabolic behavior of the yeast cells under the alpha factor or CDC28 cell cycle arrest. Under the alpha factor treatment, the central metabolic fluxes are at a high level, and the cellular metabolism tend to be respiratory even when glucose is abundant (31). The cell cycle CDK Cdc28 regulates both the cell division processes and metabolic processes. Under the CDC28 inhibition, the cells accumulate glycogen and trehalose to extremely high levels (32). Given the different characteristics of the two cell cycle arrest mechanisms, it is understandable that after the release of cell cycle arrest, the cells proceed from very different metabolic situations, and metabolism will adapt to those situations. Supporting Figure 9 shows genes associated with DC5, where we can observe a very strong pattern in the CDC28 samples, and a weaker pattern in the alpha factor samples. Functionally, we observe the highly connected biological processes mostly involve small molecule metabolism and transport (Figure 4b). Two typical pairs of genes are shown in Figure 4c, where clear dynamic correlation is observed.

Overall, unlike traditional methods such as PCA and SPCA that identify commonalities, the DCA approach tend to find signals that differentiate the four underlying experiments, and reveals some important biological processes that behave differently between the experiments. Given the existing knowledge on the dataset, these results validate that DCA extract new and meaningful information. However, in most other applications, information such as sample grouping are not available. We next examine the TCGA breast cancer (BRCA) dataset to see if the method can extract any new insights from the data.

*DCA brings new insights into the TCGA Breast Cancer data*

The data contains the measurement of 20532 genes by deep sequencing in 762 subjects with breast cancer. After removing genes with >20% zero readings, 17728 genes remain in the study. Similar to the yeast cell cycle data, the DCA captures signals that are distinct from traditional methods. Here we focus our discussion on three of the DCs, as they are clearly linked to estrogen receptor (ER) status (Figure 5a, Supporting Figure 10). DC1 largely separates ER-positive and ER-negative samples, which agrees with the second principal component very well (Figure 5b). On the other hand, in the space spanned by DC3 and DC7, ER-positive samples are tightly clustered in the middle, while part of the ER-negative samples are spread widely (Figure 5a, Supporting Figure 10). No PCs capture a similar structure in the data (Supporting Figure 11).

Further analyses show that among the ER-negative subjects, those with more extreme scores in either DC3 or DC7 show a different survival characteristic than those in the center (Figure 5c). The subjects with more extreme scores tend to have a higher chance of dying earlier, while in long follow-ups the remaining subjects tend to survive longer, albeit supported by relatively few data points.

Functionally, the biological processes that show excessive dynamic correlations conditioned on DC3 are centered around two main themes (Figure 6a). The first is protein sumoylation and stress response. Sumoylation is a post-translational modification that often occurs in response to cellular stress (33). Many oncogenes and tumor suppressors are functionally related to sumoylation (34). The second main theme is cell differentiation and tissue development that are related to several types of tissues, indicating a dysregulation in the cells. Genes associated with DC3 mainly fall into two groups that exhibit inverse correlation when DC3 score is low, and low expression when DC3 score is high (Supporting Figure 12).

The biological processes associated with DC7 are mostly immune response processes (Figure 6b). Patterns of immune cell infiltration has been linked to the prognosis and treatment response of breast cancer (35). An examination of the genes associated with DC7 reveals that over half of such genes are lowly expressed when DC7 score is more extreme. A smaller portion of the genes are lowly expressed when DC7 score is low, and highly expressed when DC7 score is high (Supporting Figure 13). In this situation, the method in fact detects a latent factor that has conditional mean-shift effects on the immune genes, which was discussed by Ho *et al* (19). The changed expression patterns of mostly immune-related genes in these samples are likely reflective of a certain immune cell infiltration pattern that has implications in prognosis. Beside the three DCs that we discuss here, most of the other DCs show clear functional implications, but require extra study beyond this manuscript to elucidate their biological meaning. The full results are in Supporting File 3.

Overall, as a new unsupervised learning method for high dimensional data, DCA can extract new and useful information from the data. It complements existing dimension reduction methods to reveal more internal structure in the data that could lead to new biological discovery. The method is straight-forward, and the computation is efficient. The R package is available at https://cran.r-project.org/web/packages/DCA/index.html.

## Acknowledgments

This work was partially supported by NIH grants U19AI090023 and U19AI057266. The author thank Mr. Yunchuan Kong, Dr. Jian Kang, and Dr. Peter Song for helpful discussions.

**Figures**

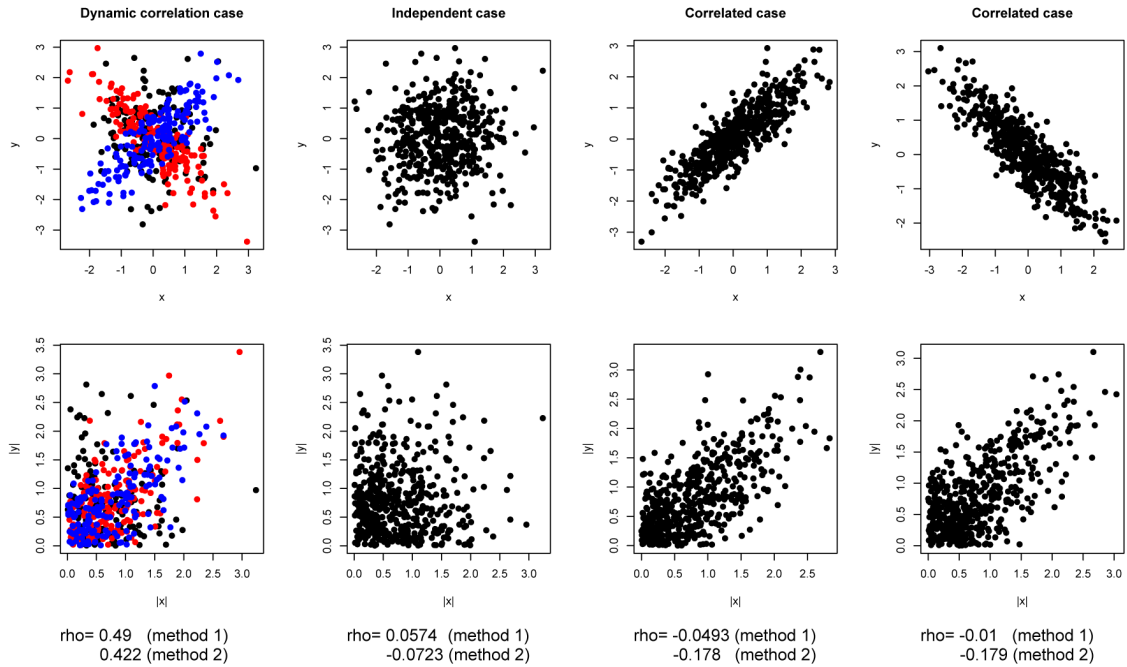

**Figure 1. Illustration of liquid association coefficient (LAC).** Left column: dynamic correlation with an unknown conditioning factor. When the factor is low, *x* and *y* are negatively correlated; when the factor is high, *x* and *y* are positively correlated. Second left column: independent case. Right two columns: correlated case. In all the cases, the marginal distribution of *X* and *Y* are standard normal.

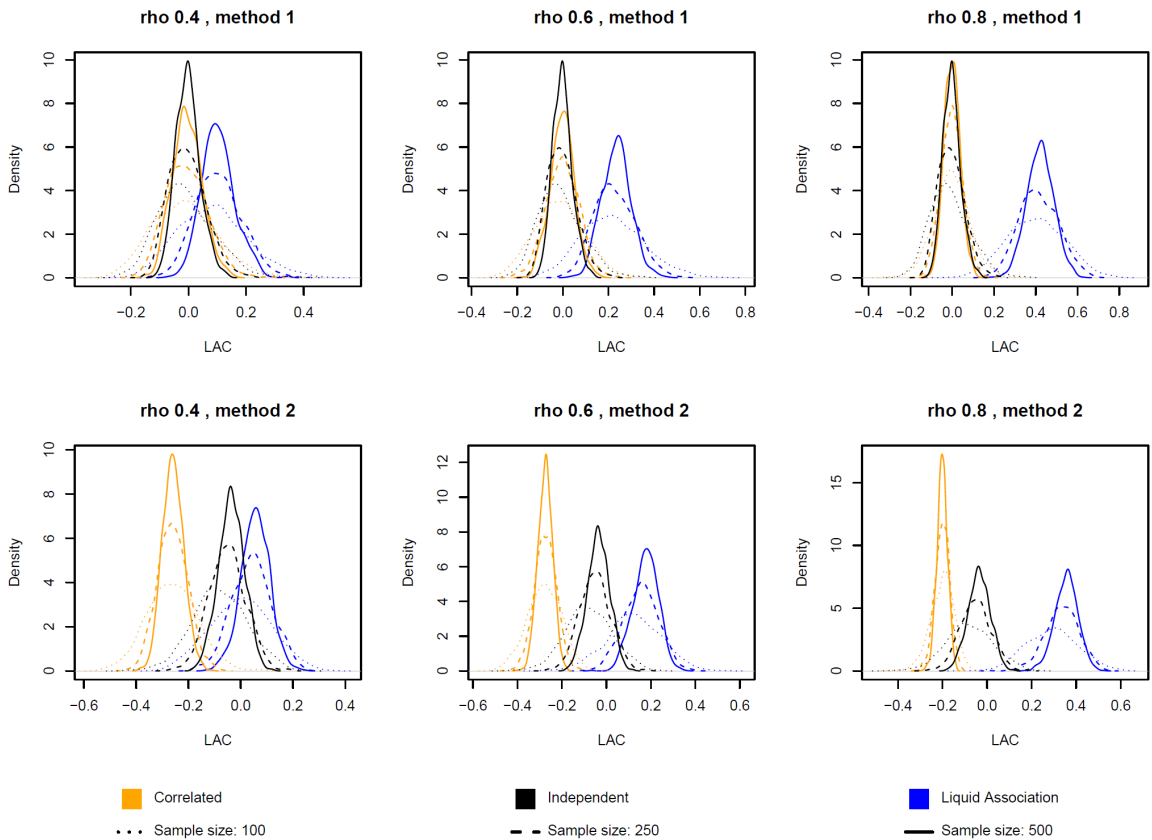

**Figure 2. Empirical distributions of LAC score under conditions of dynamic correlation, simple correlation, or independence.** The densities are based on 1000 simulations. In the dynamic correlation cases, one-third of the data points follow a bivariate normal distribution with mean $\begin{pmatrix} 0 \\ 0 \end{pmatrix}$ and variance-covariance matrix $\begin{pmatrix} 1 & \rho \\ \rho & 1 \end{pmatrix}$, one-third follow a bivariate normal distribution with mean $\begin{pmatrix} 0 \\ 0 \end{pmatrix}$ and variance-covariance matrix $\begin{pmatrix} 1 & -\rho \\ -\rho & 1 \end{pmatrix}$, and another one-third follow independent standard normal distributions. In the correlated case, all data points follow a bivariate normal distribution with mean $\begin{pmatrix} 0 \\ 0 \end{pmatrix}$ and variance-covariance matrix $\begin{pmatrix} 1 & \rho \\ \rho & 1 \end{pmatrix}$.

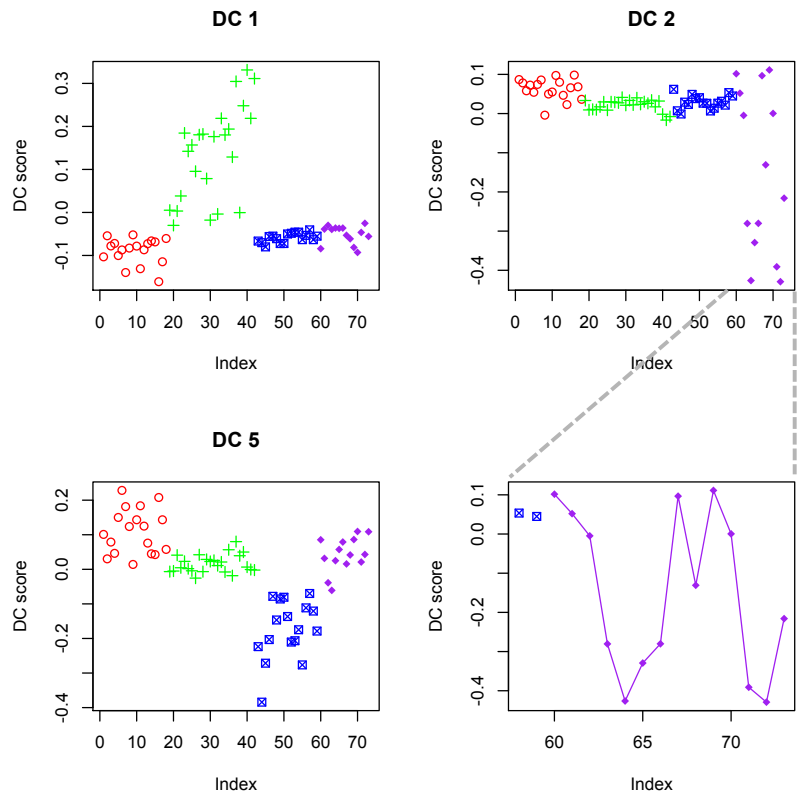

**Figure 3. Some example Dynamic Components from the cell cycle data.** Colors: the four cell cycle experiments. Red: alpha factor; green: CDC15; blue:CDC28; purple: elutriation.

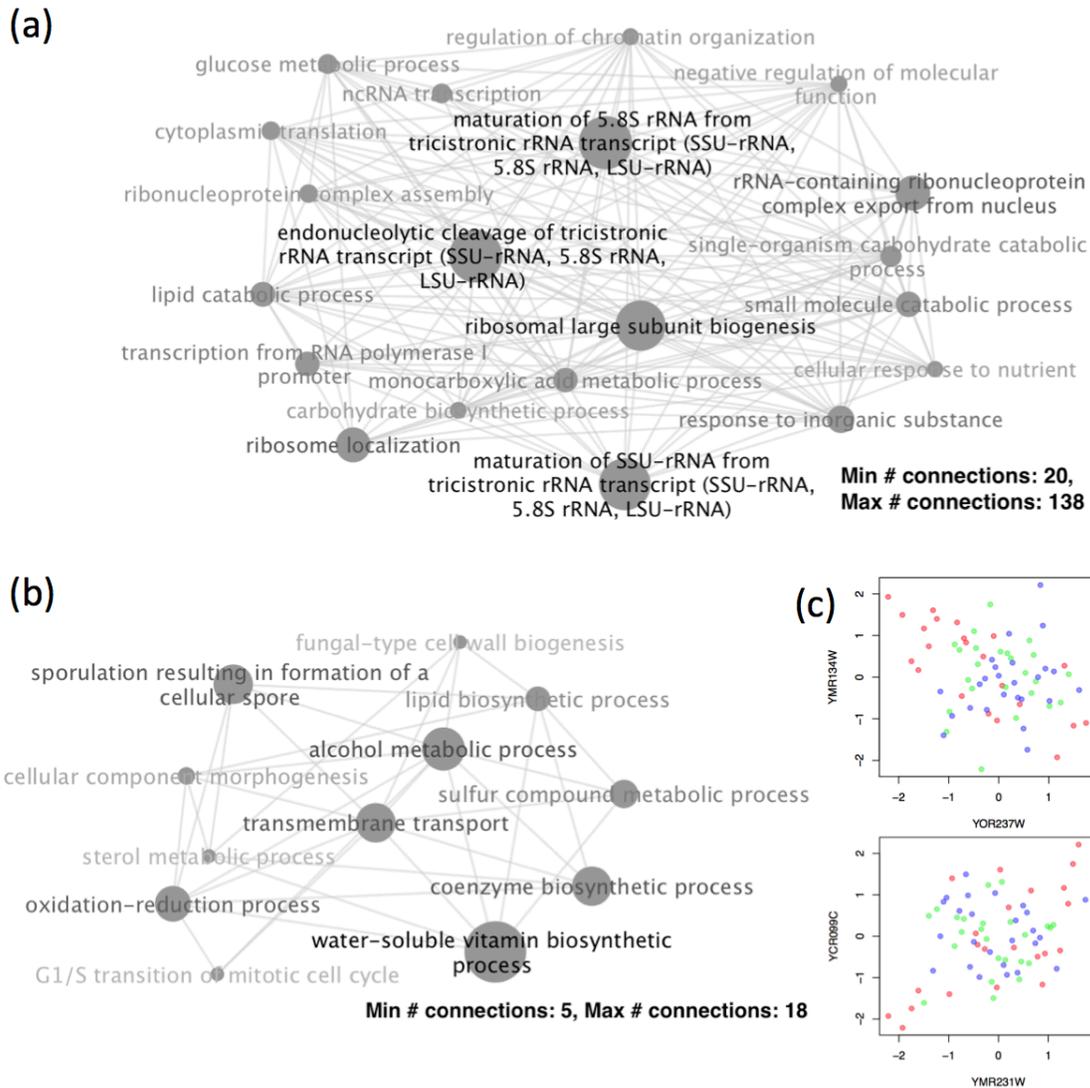

**Figure 4. Biological process pairs with excessive dynamic correlations related to DCs 2 and 5.** Gene pairs were selected using fdr threshold of 0.01. Biological process pairs were selected using a p-value threshold of 0.001 and fold-change of 2. For simplicity, only nodes with connections above a certain threshold are shown. Node sizes reflect the total number of connections of each node. (a) Biological process pairs associated with the 2nd DC. (b) Biological process pairs associated with the 5th DC. (c) Example plots of gene pairs with LA relation with DC5. Red points: samples in the lower 33% of DC5 score; blue points: samples in the upper 33% of DC5 score.

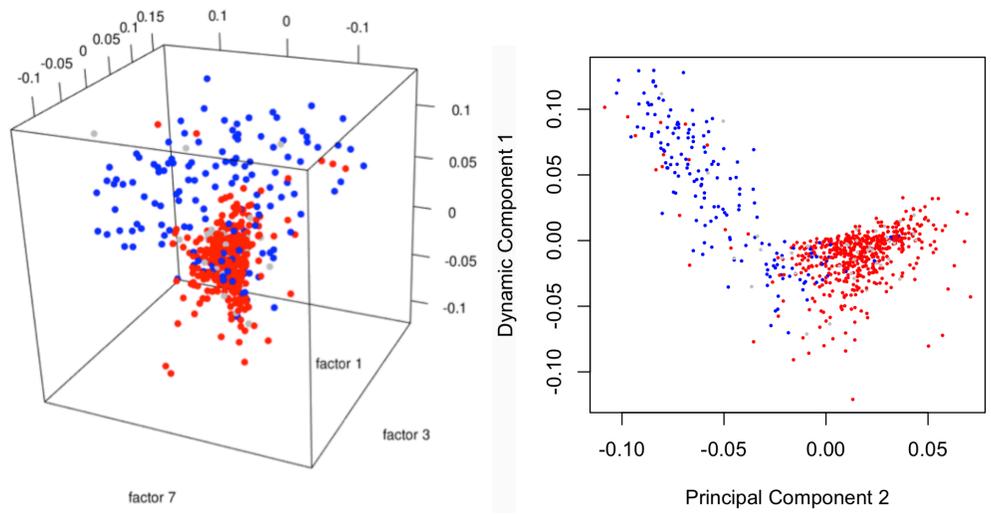
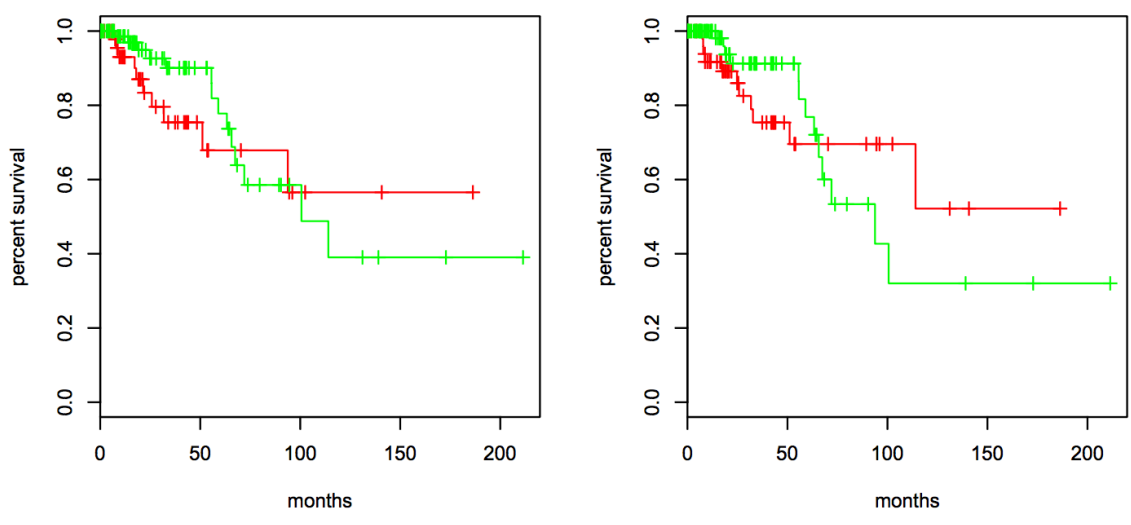

**Figure 5. Results from the TCGA BRCA dataset.** (a) Scatter plots of DC1, DC3, and DC7 scores. The points are colored based on the ER status of the subjects. DC1 separates ER+ and ER-, while DC3 and DC7 have a wide spread only for the ER- subjects. (b) DC1 captures similar information as the second principal component. (c) Survival curves of the ER-negative subjects, red: absolute factor score > 0.05.

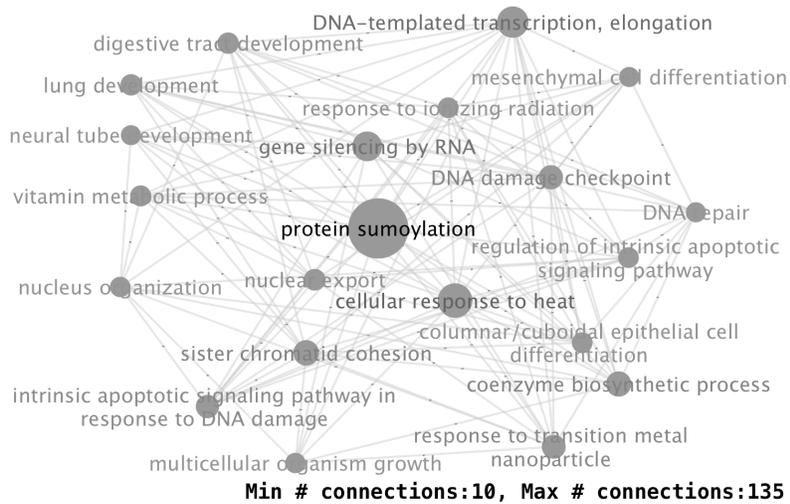

(a)

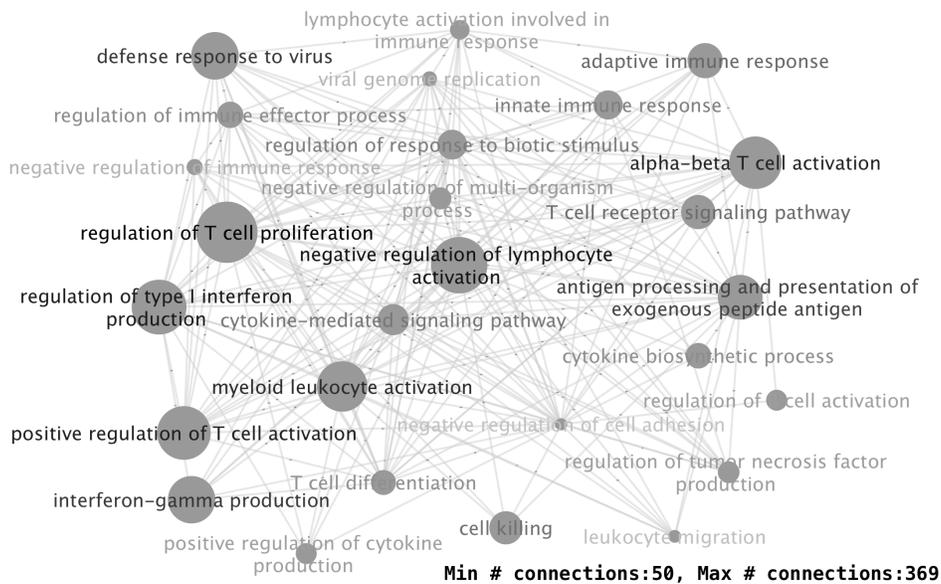

(b)

**Figure 6. Biological process pairs with excessive dynamic correlations related to DCs 3 and 7.** Gene pairs were selected using fdr threshold of 0.01. Biological process pairs were selected using a p-value threshold of 0.001 and fold-change of 3. For simplicity, only nodes with connections above a certain threshold are shown. Node sizes reflect the total number of connections of each node. (a) Biological process pairs associated with the 3rd DC. (b) Biological process pairs associated with the 7th DC.